\documentclass[twocolumn,floatfix,prl,aps,showpacs, longbibliography]{revtex4-1}
\usepackage{graphicx,amsmath,amssymb,color}
\usepackage{nicefrac,xcolor}

\newcommand{\avg}[1]{{\langle #1 \rangle}}
\newcommand{\ket}[1]{{|#1\rangle}}
\newcommand{\amp}[2]{\langle #1 | #2 \rangle}

\renewcommand{\vec}[1]{{\bf {#1}}}

\begin{document}

\title{Current-induced Gap Opening in Interacting Topological Insulator Surfaces}
\author{Ajit C. Balram$^{1,2,3}$, Karsten Flensberg$^{2}$, Jens Paaske$^{2}$, and Mark S. Rudner$^{1,2}$}
\affiliation{$^{1}$Niels Bohr International Academy, Niels Bohr Institute, University of Copenhagen, 2100 Copenhagen, Denmark}
\affiliation{$^{2}$Center for Quantum Devices, Niels Bohr Institute, University of Copenhagen, 2100 Copenhagen, Denmark}
\affiliation{$^{3}$The Institute of Mathematical Sciences, HBNI, CIT Campus, Chennai 600113, India}
\date{\today}

\begin{abstract} 
Two-dimensional topological insulators (TIs) host gapless helical edge states that are predicted to support a quantized two-terminal conductance. 
Quantization is protected by time-reversal symmetry, which forbids elastic backscattering.
Paradoxically, the current-carrying state itself breaks the time-reversal symmetry that protects it.
Here we show that the combination of electron-electron interactions and momentum-dependent spin polarization in helical edge states gives rise to feedback through which an applied current opens a gap in the edge state dispersion, thereby breaking the protection against elastic backscattering. Current-induced gap opening is manifested via a nonlinear contribution to the system's $I-V$ characteristic, which persists down to zero temperature. We discuss prospects for realizations in recently discovered large bulk band gap TIs, and an analogous current-induced gap opening mechanism for the  surface states of three-dimensional TIs.
\end{abstract}

\maketitle

Nonequilibrium many-body systems may host a variety of internal fields, such as dc currents or ac electric fields, which are not allowed in equilibrium. Through electron-electron interactions, such fields may give rise to intriguing feedback effects that lead to novel types of nonlinear transport phenomena and dynamical phase transitions~\cite{Kumai99, Dalidovich04, Green05, Feldman05,  Mitra08, Diehl10, Fausti11, Rudner18}. 
By manipulating these internal fields, we may furthermore obtain new routes for quantum engineering of material properties~\cite{Basov17}.

In this work we study how applied currents in combination with electron-electron interactions can modify the electronic properties of topological insulator (TI) surface states. In equilibrium, TIs feature a bulk band gap and topologically protected gapless Dirac-like helical surface states~\cite{Hasan10}. The degeneracy on the surface is protected by time-reversal symmetry (TRS), and is robust against weak perturbations that preserve this symmetry. 
Here we propose a novel gap opening mechanism that is important for characterizing the breakdown of topological protection due to the unavoidable TRS breaking that accompanies an applied current. 

Previous works have explored possibilities of spontaneous breakdown of topological surface states in equilibrium due to electron-electron interactions~\cite{Wu2006, Xu2006, Baum12, Black-Schaffer14, Chou18, Chou19}, as well as means for opening a gap on the surface by explicitly breaking TRS via magnetic doping~\cite{Liu08, Chen10, Yu10, Zhu11, Fox18}, coupling to the exchange field of an adjacent magnetic layer~\cite{Tang17} or the nuclear spins of the host material~\cite{Maestro13, Hsu17, Hsu18}, or through the application of a magnetic field~\cite{Konig07, Dominguez18, Skolasinski18}, or circularly polarized light~\cite{Wang13}.

\begin{figure}[t]
\begin{center}
\includegraphics[width=\columnwidth]{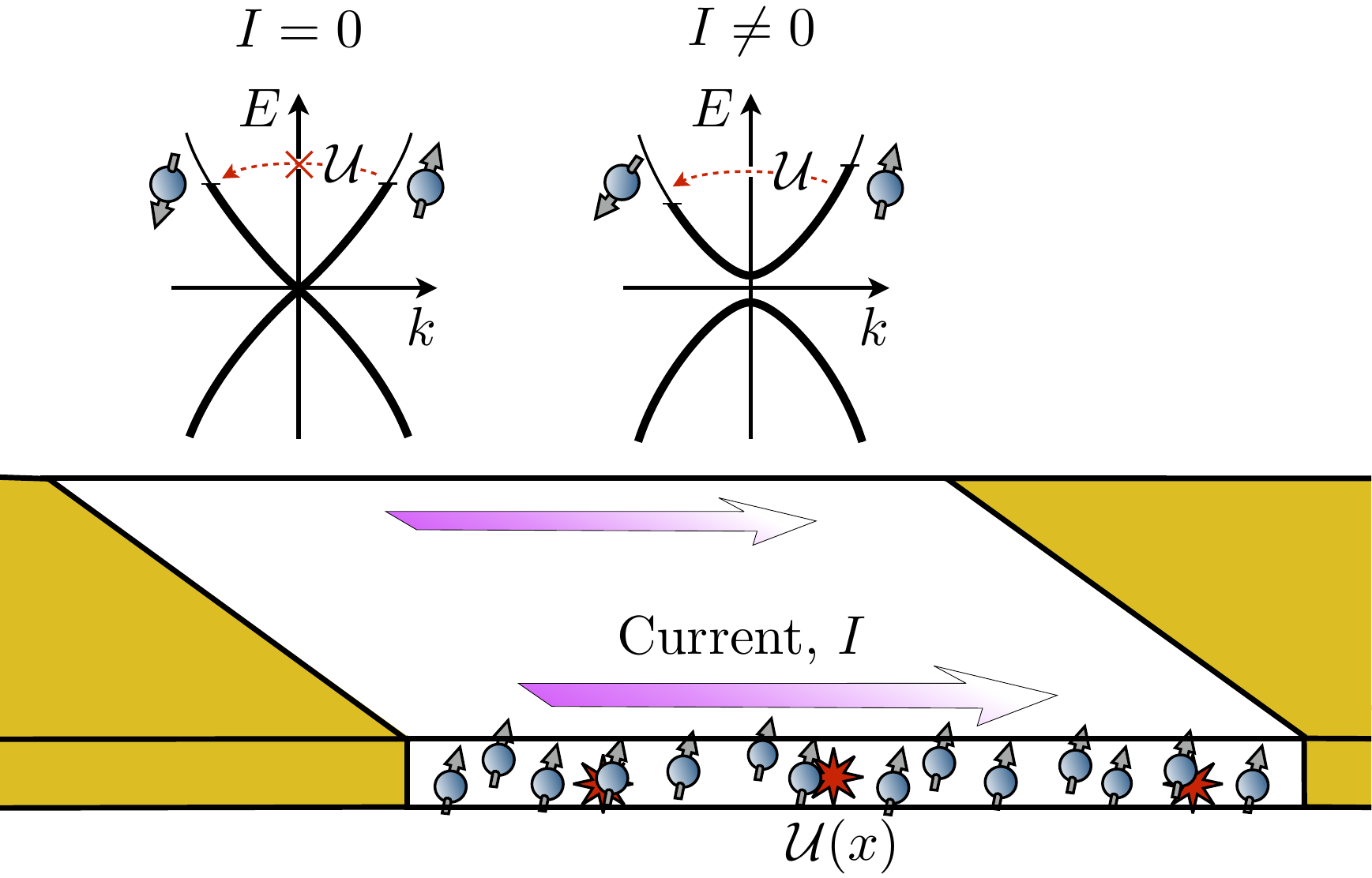}
\end{center}
\vspace{-0.1in}
\caption{Current-induced gap opening on the edge of an interacting two-dimensional topological insulator.
In equilibrium, the edge hosts a one-dimensional gapless helical mode (upper left). 
When a current is driven through the system, a net spin polarization develops due to spin-orbit coupling (upper right, bottom). Due to electron-electron interactions, the spin polarization induces a nonvanishing internal exchange field, which acts back via the momentum-dependent spin quantization axis [Eq.~(\ref{eq_Hamiltonian_QSH_edge})] to open a gap in the helical dispersion (upper right). Once Kramers's degeneracy is lifted, the edge mode is no longer protected against elastic backscattering by a disorder potential, $\mathcal{U}(x)$ (red stars). }
\label{fig:intro}
\vspace{-0.2in}
\end{figure}

Importantly, in the presence of a dc current, the system lacks TRS. Therefore, by symmetry considerations, an applied current can break the degeneracy and open a gap in the helical surface state spectrum. However, in the absence of electron-electron interactions, the applied current simply results from a nonequilibrium population of the single particle surface states, and there is no feedback mechanism through which the current may modify the spectrum itself. Crucially, due to the spin-orbit coupling that is naturally present in TI materials, the applied current also carries a spin polarization~\cite{Edelstein90,Brune12}. Through electron-electron interactions, the spin polarization generates an {\it internal exchange field} that can open a gap in the surface state dispersion (see Fig.~\ref{fig:intro}).

The gap opening mechanism that we discuss is general, and can be applied to the surface states of both two-dimensional (2D) and three-dimensional (3D) TIs. The nontrivial portion of the feedback, which leads to a gap, is mediated through the part of the spin-orbit coupling that causes the spin helicity axis to rotate as a function of energy~\cite{Schmidt12, Rod15, Ortiz16}. We first illustrate the mechanism of current-induced gap opening for the case of the one-dimensional helical edge states of a 2D TI. We investigate the dependence of backscattering due to nonmagnetic impurities on the applied current, and discuss the resulting nonlinear current-voltage characteristic. We then discuss the extension of this mechanism to the 2D surface states of 3D TIs.

A generic one-dimensional (1D) helical mode on the edge of a 2D TI is characterized by a dispersion relation and a spin helicity axis that determines the directions of the spinor eigenstates for right ($r$) and left ($l$) movers as a function of the electronic wave number (momentum) $k$ parallel to the edge~\cite{Schmidt12}.
Time reversal symmetry imposes the constraints that the energy of the left mover at momentum $-k$ must be equal to the energy of the right mover at momentum $k$, and that their associated spins must be opposite. The latter condition ensures that elastic backscattering by nonmagnetic impurities is forbidden due to the orthogonality of the initial and final spin states~\cite{footnote:no_backscattering}. Generically, the spin helicity axes at different values of $k$ are not parallel~\cite{Schmidt12, Rod15, Ortiz16}. 

\begin{figure}[t]
\begin{center}
\includegraphics[width=\columnwidth]{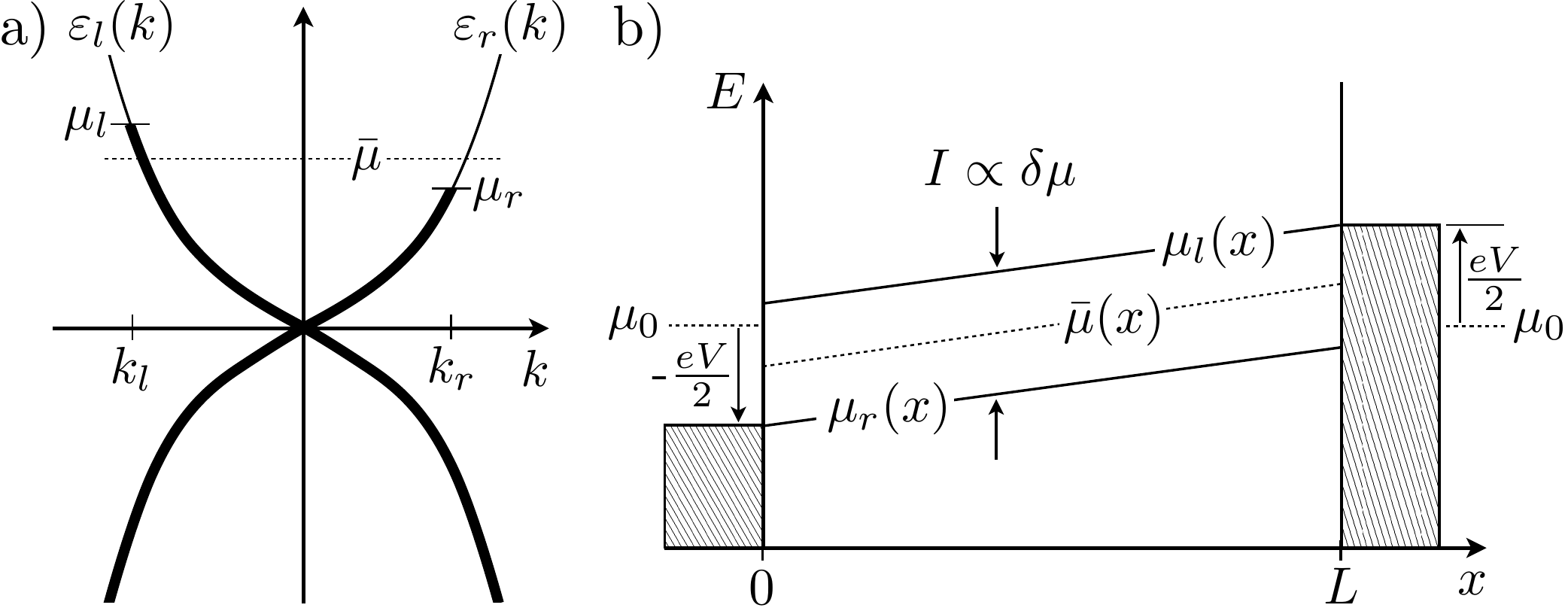}
\end{center}
\caption{Current-induced gap opening and nonlinear resistance in 1D helical edge modes of a 2D TI. 
a) Unperturbed edge mode dispersion [see Eq.~(\ref{eq_Hamiltonian_QSH_edge})]. In equilibrium, the system possesses a finite chemical potential $\bar{\mu} = \mu_{0}$ above the Dirac point. We describe the current-carrying state by imposing a (local) difference between chemical potentials $\mu_{r}$ and $\mu_{l}$ of the right- and left-moving modes, respectively. The corresponding Fermi momenta are labeled $k_{r}$ and $k_{l}$. b) Setup for calculating the nonlinear $I-V$ characteristic, Eq.~(\ref{eq:IV}). A segment of helical edge state is contacted by two reservoirs with chemical potentials $\mu_{0} - eV/2$ and $\mu_{0} + eV/2$, at $x = 0$ and $x = L$, respectively. The chemical potential of the left [right] reservoir sets the boundary condition for $\mu_{r}(x)$ [$\mu_{l}(x)$] at $x = 0$ [$x = L$]. The current $I$ is proportional to $\delta \mu \equiv \mu_{l}(x) - \mu_{r}(x)$, which is independent of $x$ due to current conservation.}
\label{fig:1D}
\end{figure}

To illustrate current-induced gap opening, we consider a minimal model that exhibits the necessary ingredients of (i) TRS, (ii) a $k$-dependent spin-helicity axis, and (iii) electron-electron interactions.
We focus our attention on an extended single edge of a 2D TI, assuming that the sample is wide enough to prohibit interedge scattering (cf.~\cite{Chang11}). At the single-particle level, the system is described by the Hamiltonian~\cite{Schmidt12, Rod15, Ortiz16, footnote:k3sz}:
\begin{equation}
 H_{1D}(k)=\hbar vk\sigma_{z} + \lambda k^{3} \sigma_{x}, 
 \label{eq_Hamiltonian_QSH_edge}
\end{equation}
where $v$ is the velocity at small $k$, and $\boldsymbol{\sigma} = (\sigma_x, \sigma_y, \sigma_z)$ is the vector of Pauli matrices describing the electron spin. 
The parameter $\lambda$ controls the rate of rotation of the spin helicity axis as a function of $k$. 
The edge mode dispersion relation corresponding to Eq.~(\ref{eq_Hamiltonian_QSH_edge}) is given by $\varepsilon_r(k)= {\rm sign}(k)\sqrt{(\hbar vk)^{2} + (\lambda k^{3})^2}$, $\varepsilon_{l}(-k) = \varepsilon_{r}(k)$ [see Fig.~\ref{fig:1D}a)]. 
At zero temperature, the states are filled up to the chemical potential, $\mu_{0}$, which we take to be greater than zero (without loss of generality).

Current-induced gap opening arises via the spin polarization that naturally accompanies a net current in the system, due to the spin-orbit coupling in Eq.~(\ref{eq_Hamiltonian_QSH_edge}). Consider a homogeneous, nonequilibrium setting in which a current $I$ flows due to an imbalance of chemical potentials $\mu_{l}$ and $\mu_{r}$ in the left- and right-moving modes, respectively [see Fig.~\ref{fig:1D}a)]: $\mu_{r} = \bar{\mu} - \delta \mu/2$, $\mu_{l} = \bar{\mu} + \delta \mu/2$, where $\bar{\mu}$ is the average chemical potential on the edge.  The current $I$ is proportional to $\delta \mu$. We will work in the limit where $\lambda \ll \hbar^{3}v^{3}/\bar{\mu}^2$ and where $\delta \mu \ll \bar{\mu}$, for simplicity. In this regime, we obtain the spin polarization and induced gap analytically, to linear order in $\lambda$ and $\delta \mu$.

Summing up the spin polarizations of all filled states between the left and right Fermi wave vectors $k_{l}$ and $k_{r}$, respectively [see Fig.~\ref{fig:1D}a)], we obtain the current-induced spin density $\avg{\vec{s}} = \int_{k_{l}}^{k_{r}}\frac{dk}{2\pi} \langle\psi_{k+}|\frac{\boldsymbol{\hbar\sigma}}{2}|\psi_{k+}\rangle$, where $\ket{\psi_{k+}}$ is the eigenvector of $H_{1D}$ in Eq.~(\ref{eq_Hamiltonian_QSH_edge}) with positive energy. Expanding $k_{r}$ and $k_{l}$ to linear order in $\lambda$ and $\delta\mu$, $k_{r}=\frac{\bar{\mu}}{\hbar v}-\frac{\delta\mu}{2\hbar v}+\mathcal{O}(\lambda^{2})$, $k_{l}=-\frac{\bar{\mu}}{\hbar v}-\frac{\delta\mu}{2\hbar v}+\mathcal{O}(\lambda^{2}) $, we find~\cite{footnote:self_consistency}:
\begin{equation}
  \avg{s_x} \approx -\frac{\lambda \bar{\mu}^{2} }{\hbar^{3}v^{3}}\frac{\delta\mu}{4\pi v},\ \avg{s_y} = 0,\ \avg{s_z} \approx -\frac{\delta\mu}{4\pi v}.
  \label{eq_spin_after_I}
\end{equation}

At mean field (Hartree-Fock) level, the spin polarization (\ref{eq_spin_after_I}) produces a Zeeman-like internal exchange field through the electron-electron interaction. For illustration we consider a contact interaction potential $U(R) = g \delta(R)$, where $g$ is the interaction strength and $R$ is the interparticle distance. The exchange field gives rise to the mean-field interaction Hamiltonian: $H^{\rm MF}_{\rm int}=-g \avg{\vec{s}} \cdot \boldsymbol{\sigma}/\hbar $. Combining $H^{\rm MF}_{\rm int}$ with $H_{1D}(k)$, Eq.~(\ref{eq_Hamiltonian_QSH_edge}), we obtain the mean-field Hamiltonian: 
\begin{eqnarray}
  H_{1D}^{\rm MF} = (\hbar vk -g\avg{s_z}/\hbar )\sigma_{z} + 
  (\lambda k^{3} - g\avg{s_x}/\hbar)\sigma_{x}.
  \label{eq_Hamiltonian_QSH_edge_cubic_SOC_driven_interacting}
\end{eqnarray}

For nonzero $\avg{s_x}$ or $\avg{s_z}$, the Hamiltonian in Eq.~(\ref{eq_Hamiltonian_QSH_edge_cubic_SOC_driven_interacting}) lacks TRS. Importantly, however, for $\lambda = 0$, the term proportional to $\sigma_x$ in Eq.~(\ref{eq_Hamiltonian_QSH_edge_cubic_SOC_driven_interacting}) vanishes and the edge state remains gapless [with the Dirac point shifted to $k_* = g\avg{s_z}/(\hbar^{2}v)$]. 
The energy-dependent spin helicity rotation is essential for gap opening, as it introduces a non-commuting term in the Hamiltonian. 

The Hamiltonian $H^{\rm MF}_{1D}$ gives rise to the dispersion $\varepsilon^{\rm MF}_\pm(k) = \pm \sqrt{(\hbar vk - g\avg{s_z}/\hbar)^2 + (\lambda k^3 -g\avg{s_x}/\hbar)^2}$. 
(Here $+$ and $-$ refer to the positive and negative energy bands.)
We seek the (shifted) position of the upper band minimum, $k_*$, and the magnitude of the induced gap, $\Delta$, by setting $\frac{d}{dk}\varepsilon^{\rm MF}_+(k)\vert_{k_*} = 0$. This condition yields a fifth-order polynomial for $k_*$: $3 \lambda^2 k_*^5 - 3 \lambda g\avg{s_x} k_*^2/\hbar + \hbar^2 v^2 k_* - vg\avg{s_z} = 0$. In the absence of a current, $\delta \mu = 0$, the Dirac point resides at $k_* = 0$.  For small currents, characterized by the limit $\delta \mu \ll 4 \pi \frac{\hbar v}{g} \bar{\mu}$, and in the small $\lambda$ limit taken above, the shift of $k_*$ is small and the terms beyond linear order in $k_*$ in the polynomial above can be neglected. Within this regime, to leading order in $\delta \mu$, we find $k_* \approx g\avg{s_z}/(\hbar^{2}v) = -\frac{1}{4\pi}\frac{g}{\hbar v} \frac{\delta \mu}{\hbar v}$.
Evaluating the dispersion at $k_*$, we obtain the induced gap: 
\begin{equation}
\label{eq:gap1D} \Delta \approx 2\left|\frac{g\avg{s_x}}{\hbar}\right| = \frac{1}{2\pi}\left|\frac{g}{\hbar v}\frac{\lambda \bar{\mu}^2}{\hbar^{3}v^3} \delta \mu\right|.
\end{equation}
Note that the assumptions on small $\delta \mu$ and $\lambda$ were taken above only to facilitate obtaining a simple analytical result; the qualitative results do not rely on this limit~\cite{footnote:velocity}.

The induced gap can be observed directly via tunneling spectroscopy, or indirectly through its nonperturbative effects on transport. Due to the fact that $H^{\rm MF}_{1D}$ lacks TRS, the protection against elastic backscattering is lost: even a nonmagnetic disorder potential $\mathcal{U}(x)$ couples states with opposite helicities, yielding dissipation on the edge (see Fig.~\ref{fig:intro}). 

To investigate the nonlinear current-voltage characteristic for a helical edge mode with a current-induced gap,  we consider an edge of length $L$, contacted by Fermi reservoirs at $x = 0$ and $x = L$ with chemical potentials $\mu_{0} - eV/2$ and $\mu_{0} + eV/2$, respectively [see Fig.~\ref{fig:1D}b)]. Here $-e < 0$ is the electron charge and $V$ is the bias voltage.

We consider the situation where forward scattering leads to rapid local equilibration separately within the left- and right-moving branches. Near the Fermi surface, the edge mode populations $f_{l}(k; x)$ and $f_{r}(k; x)$ can be described by Fermi-Dirac distributions with separate, position-dependent chemical potentials $\mu_{l}(x)$ and $\mu_{r}(x)$, respectively. The chemical potential of the left reservoir (at $x = 0$) provides a boundary condition for the chemical potential of the right movers, $\mu_{r}(x = 0) = \mu_{0} - eV/2$; similarly the right reservoir provides a boundary condition for the left movers: $\mu_{l}(x = L) = \mu_{0} + eV/2$.

Elastic backscattering, leading to resistance, is induced by the disorder potential $\mathcal{U}(x)$ in the presence of the current-induced gap. We assume that $\mu_{l}(x)$ and $\mu_{r}(x)$ vary gradually in space such that, locally (at each position, $x$), the dispersion and scattering properties of the system are described by Hamiltonian (\ref{eq_Hamiltonian_QSH_edge_cubic_SOC_driven_interacting}), with the local value of the spin polarization $\avg{\vec{s}}(x)$ determined by Eq.~(\ref{eq_spin_after_I}) with $\bar{\mu}$ replaced by $\bar{\mu}(x) =[\mu_{l}(x) + \mu_{r}(x)]/2$.

To leading order in $\lambda$ and $\delta \mu$, the matrix element for disorder-induced elastic scattering of a right mover with momentum near $k_{r}$ to a left mover with momentum near $k_{l}$ is found to be proportional to the overlap $\amp{\psi^{\rm MF}_{k_{l}+}}{\psi^{\rm MF}_{k_{r} +}} \approx  \frac{g\avg{s_x}}{\hbar \mu}$. Here $\mu \approx \bar{\mu}(x)$ is the energy of the scattered electron, and $\ket{\psi^{\rm MF}_{k+}}$ is the positive energy eigenvector of $H^{\rm MF}_{1D}$, Eq.~(\ref{eq_Hamiltonian_QSH_edge_cubic_SOC_driven_interacting}). 
We consider short-range correlated disorder, characterized by $\overline{\mathcal{U}(x)\mathcal{U}(x')} = \frac{(\hbar v)^2}{\ell}\delta(x - x')$, where the overline indicates averaging over disorder and we express the disorder strength through the effective mean free path, $\ell$, that would arise in a comparable {\it nonhelical} 1D system. 
Within this model, and in the limit $k_r \ell, k_l \ell \gg 1$, we obtain the disorder-averaged backscattering rate $1/\tau$ perturbatively from Fermi's golden rule: 
\begin{equation}
\frac{1}{\tau} = 
\frac{v}{\ell}\left(\frac{g\lambda\bar{\mu}(x)\delta\mu}{4\pi\hbar^4v^4}\right)^2,
\label{eq:def_tau_independent_L}
\end{equation}
where we used $1/(2\pi\hbar v)$ as the density of states per unit length (valid for small $\lambda$ and small bias~\cite{footnote:velocity}). Due to the current-induced nature of the gap, $\tau^{-1}$ is proportional to $(\delta\mu)^2$ and hence to the square of the applied current.

We seek the steady-state current that flows as a function of the bias voltage, $V$. To this end, we study the left- and right-mover charge densities, $\rho_{d}(x) = (-e)\int \frac{dk}{2\pi}\, f_{d}(k; x)$, with $d = \{r, l\}$, and corresponding currents, $I_{r}(x) = v \rho_{r}(x)$, $I_{l}(x) = -v \rho_{l}(x)$. (Here we again use the approximation that the velocity is constant, valid in the limit of small $\lambda$ and bias~\cite{footnote:velocity}.) 
At zero temperature, the total current $I = I_{r} + I_{l}$ is proportional to the (local) chemical potential difference: $I = \frac{(-e)}{h}(\mu_{r} - \mu_{l}) = \frac{e}{h}\delta\mu$. Hence, conservation of total current implies that $\delta\mu \equiv \mu_{l}(x) - \mu_{r}(x) = {\rm constant}$.

To obtain the spatial profiles of $\mu_{r}(x)$ and $\mu_{l}(x)$, we use the continuity equation: $\partial_x I_{r} = - \partial_t \rho_{r}|_{\rm scatt}$, where $\partial_t \rho_{r}|_{\rm scatt} = -(\rho_{r} - \rho_{l})/\tau$ is the net rate for right movers to be scattered into the left-moving branch, and $1/\tau$ is the backscattering rate, Eq.~(\ref{eq:def_tau_independent_L}). A similar continuity equation holds for the left movers. Expressing $\rho_{r}(x)$, $\rho_{l}(x)$, $I_{r}(x)$, and $I_{l}(x)$ in terms of $\mu_{r}(x)$ and $\mu_{l}(x)$, we obtain a differential equation for $\bar{\mu}(x)$:
\begin{equation}
\label{eq:continuity_mu}\frac{\partial \bar{\mu}(x)}{\partial x} =\frac{\delta \mu}{v\tau} \equiv \mathcal{C}\bar{\mu}(x)^2 (\delta \mu)^3. 
\end{equation}
To clearly expose the  $\bar{\mu}(x)$ and $\delta \mu$ dependence on the right hand side of Eq.~(\ref{eq:continuity_mu}), we have gathered the remaining factors into the constant $\mathcal{C} = g^2\lambda^2/[\ell(4\pi\hbar^4v^4)^2]$. 

Noting that $\delta \mu$ is constant (see above), Eq.~(\ref{eq:continuity_mu}) can be integrated directly to obtain:
\begin{equation}
  \label{eq:mu_soln}\frac{1}{\bar{\mu}(L)} - \frac{1}{\bar{\mu}(0)} =  -L\mathcal{C}(\delta\mu)^3.
\end{equation}
We now solve for $\delta \mu$ (and hence the total current, $I$) as a function of the bias $eV$ by applying the boundary conditions above (see also Fig.~\ref{fig:1D}b): $\bar{\mu}(0) = \mu_{0} - \frac{1}{2}(eV - \delta\mu)$, $\bar{\mu}(L) = \mu_{0} + \frac{1}{2} (eV - \delta\mu)$. 
Substituting these expressions into Eq.~(\ref{eq:mu_soln}), we obtain a nonlinear equation for $\delta \mu$:
\begin{equation}
  \label{eq:delta_mu} eV - \delta \mu = L\mathcal{C}\mu_0^2(\delta \mu )^{3}\left[1 - \frac{1}{4}\left(\frac{eV- \delta \mu}{\mu_0}\right)^2\right] .
\end{equation}
The solutions of Eq.~(\ref{eq:delta_mu}) together with $I = \frac{e}{h}\delta\mu$ yield the $I-V$ characteristic of the system. 
Note that we have assumed $eV/\mu_0,\, \delta \mu/\mu_0 \ll 1$ throughout. 

We analyze Eq.~(\ref{eq:delta_mu}) in two limits: a ``nearly ballistic regime,'' realized for small biases and/or for short, weakly disordered systems, and a ``resistive regime,'' realized for large biases and/or in long, strongly disordered systems.
In the nearly ballistic regime, characterized by $\delta \mu \approx eV$, we let $\delta \mu = (1 - x)eV$ and solve Eq.~(\ref{eq:delta_mu}) to leading order in $x \ll 1$. 
This solution is consistent for $\kappa \equiv L\mathcal{C}\mu_{0}^2 (eV)^2 \ll 1$. 
In the resistive regime, realized for $\kappa \gg 1$ and characterized by $\delta \mu \ll eV$, we let $\delta \mu = x eV$ and again solve Eq.~(\ref{eq:delta_mu}) to leading order in $x$.  
We find:
\begin{eqnarray}
\label{eq:IV}I = \frac{e^{2}}{h}\! \cdot\! \left\{\!\! \begin{array}{rl}
 V-L\mathcal{C}\mu_{0}^{2}e^{2}V^{3}, & \!  \kappa \ll 1\ \textrm{(nearly ballistic)}\\
\left(e^{2}L\mathcal{C}\mu_{0}^{2} \right)^{-\frac13}V^{\frac13}, &\! \kappa \gg 1\ \textrm{(resistive)}.
\end{array}\right.
\end{eqnarray}
The behavior in both regimes can be understood in terms of a series resistor network consisting of a fixed contact resistance $R_0 = h/e^2$ and a nonlinear resistance $R_{\rm scatt} \sim I^2$ associated with backscattering that depends quadratically on the current, $I$ [see text below Eq.~(\ref{eq:def_tau_independent_L})]. 
In the nearly ballistic regime, the contact resistance dominates: $R_{\rm scatt}/R_0 \ll 1$. Here we may make the replacement $R_{\rm scatt} \sim V^2$, to obtain $I = V/(R_0 + R_{\rm scatt}) \sim \frac{e^2}{h}V - \mathcal{O}(V^3)$. In the resistive regime, $R_{\rm scatt}/R_0 \gg 1$, we may neglect $R_0$ to obtain $I^3 \sim V$.

The nonlinear $I-V$ characteristic (\ref{eq:IV}) should be contrasted with other nonlinear contributions to the change in conductance~\cite{Schmidt12, Xu2006, Wu2006, Vayrynen2013, Maestro13, Hsu17, Hsu18, Bagrov2018, Novelli2019}. For example, in Ref.~\cite{Schmidt12}, Schmidt \emph{et al.}~considered inelastic two-particle backscattering due to weak interactions and an impurity potential. The mechanism they considered would give a quadratic-in-voltage correction to the (nonlinear) conductance that scales with the temperature squared; the mechanism we describe yields a quadratic-in-voltage correction to the nonlinear conductance that persists down to zero temperature. Thus backscattering due to current-induced gap opening can be distinguished from other mechanisms, such as inelastic scattering, through the temperature dependence of nonlinear transport. 

The current-induced gap opening mechanism described above straightforwardly generalizes to the case of 2D surface states of a 3D TI. For illustration we take a  minimal model Hamiltonian that approximately describes the surface states of Bi$_{2}$Te$_{3}$ (see Refs.~\cite{Fu09, Lee09}):
\begin{equation}
H_{2D}(\vec{k})=\hbar v(\vec{k} \times \boldsymbol{\sigma})\cdot\hat{z}+\lambda k^{3} \cos(3\phi_{\vec{k}}) \sigma_{z},
\label{eq_Hamiltonian}
\end{equation}
where $\vec{k}=(k_{x},k_{y})$, $k=|\vec{k}|$, and $\phi_{\vec{k}}$ is the azimuthal angle of $\vec{k}$. The hexagonal warping term (with strength $\lambda$) plays a crucial role in enabling current-induced gap opening. As above, we consider the situation where the system is doped up to a finite chemical potential $\mu_0$ above the Dirac point, in the limit of temperature much less than $\mu_0$. 

In equilibrium, time-reversal symmetry of $H_{2D}$ implies that the net spin polarization vanishes, $\avg{\vec{s}} = 0$. When an electric field is applied to drive a dc current on the surface, the Fermi surface shifts and becomes distorted, yielding a finite spin polarization $\avg{\vec{s}} = \int \frac{d^{2}\vec{k}}{(2\pi)^2}\, f_{\vec{k},+} \langle\psi_{\vec{k},+}|\frac{\hbar\boldsymbol{\sigma}}{2}|\psi_{\vec{k},+}\rangle$, where $\ket{\psi_{\vec{k},+}}$ is the positive energy eigenstate of $H_{2D}(\vec{k})$, and $f_{\vec{k},+}$ is the nonequilibrium distribution function. 
Through electron-electron interactions, this spin polarization yields an internal exchange field captured by the mean field Hamiltonian: 
\begin{eqnarray}
\nonumber H^{\rm MF}_{\rm 2D}(\vec{k}) &=& (-vk_{y} - g \avg{s_{x}}/\hbar)\sigma_{x}+(vk_{x} - g \avg{s_{y}}/\hbar) \sigma_{y}\\
&&+\ \left[\lambda k_{x}(k^{2}_{x}-3k^{2}_{y}) - g\avg{s_{z}}/\hbar \right] \sigma_{z}.
\label{eq_Hamiltonian_after_E_with_interactions}
\end{eqnarray}
For $\lambda = 0$, the spin polarization is purely in plane, $\avg{s_z} = 0$, and the exchange field merely shifts the Dirac point (without opening a gap). With $\lambda \neq 0$, however, the induced spin polarization generically opens a gap in the spectrum of $H^{\rm MF}_{\rm 2D}(\vec{k})$. 

The nonequilibrium distribution function $f_{\vec{k},+}$, and hence the induced gap, can be obtained, e.g., using a Boltzmann equation approach analogous to that in Ref.~\cite{Sodemann15}. Due to the reduced rotational symmetry of the surface states described by Eq.~(\ref{eq_Hamiltonian}), the induced gap is sensitive to the direction of the applied field. This model still displays a high degree of symmetry, which in particular yields $\avg{s_z} = 0$ within the simplest Boltzmann equation treatment. Consequently, the leading contribution to the induced gap is third order in both the magnitude of the applied electric field and the interaction strength, and therefore turns out to be small for this model.

{\it Discussion.---}
We have analyzed a new mechanism through which the interplay between nonequilibrium currents and electron-electron interactions can modify the properties of TI surface states. For the 1D edge of a 2D TI, the magnitude of the induced gap (within the limits of small $\lambda$ and $\delta \mu$ taken throughout) is given by Eq.~(\ref{eq:gap1D}). Due to the scaling with $\bar{\mu}^2$ and $1/v^4$, a current-induced gap opening is expected to be the most pronounced in topological insulators with large bulk gaps (allowing for large $\bar{\mu}$ above the Dirac point), and low edge state velocities.
Bulk gaps of the order of $100~{\rm meV}$ and above have been predicted in two-dimensional transition metal dichalcogenides such as WTe$_2$~\cite{Qian14, Fei17, Wu18}, as well as, e.g., jacutingaite~\cite{Marrazzo18} and (functionalized) stanene (tin) films~\cite{Xu13}. To estimate the induced gap, we take $\bar{\mu}=100~{\rm meV}$, $v= 10^{5}$ m/s~\cite{Qian14}, and $g = e^2/(4\pi\epsilon_{r}\epsilon_{0}) \approx 2.4~{\rm eV\, \AA}$~\cite{footnote:g_value}, where we use $\epsilon_{r} \approx 6$ as the relative dielectric constant characterizing the 2D TI and its surroundings~\cite{Laturia18} ($\epsilon_{0}$ is the vacuum permittivity). Equation~(\ref{eq:gap1D}) then gives $\Delta \approx 0.02\, \lambda [{\rm eV\, \AA}^3]\, \delta \mu$. The value of $\lambda$ is currently unknown for the large bulk-gap materials mentioned above; however, based on trends in the {\it ab initio} data of Ref.~\cite{Rod15}, for WTe$_2$ we speculate that it may fall in the range $\lambda \sim 10^{-2} - 10^{-1}$ eV \AA$^3$, leading to a gap of the order of $1-10$ $\mu$eV for a bias voltage of a few mV. In the nearly ballistic regime of Eq.~(\ref{eq:IV}), the resulting fractional change to the ballistic conductance $e^2/h$ is given by $L \mathcal{C}\mu_0^2 (eV)^2 \approx \frac{L}{\ell} \left(\frac{\Delta}{2\mu_0}\right)^{2}$.

For the 2D surface states of 3D TIs, current-induced gap opening is accompanied by the appearance of a Berry curvature monopole that gives rise to an anomalous Hall effect. Interestingly, this nonlinear Hall effect is of a fundamentally different origin than other nonlinear anomalous Hall effects that have been the subject of intense recent interest~\cite{Sodemann15, Ma18, Xu18,You18, Kang18, Konig18}. While these works investigate nonlinear transport due to a Berry curvature {\it dipole} that is present in the equilibrium band structure of the system, the mechanism we describe arises from the Berry curvature {\it monopole} that arises due self-modification of the system's nonequilibrium band structure via electron-electron interactions. The search for platforms where this nonlinear Hall effect can be enhanced is an interesting direction for future work.

\begin{acknowledgments}
We thank D.~Pesin and A.~Stern for stimulating discussions. The Center for Quantum Devices is funded by the Danish National Research Foundation. This work was supported by the European Research Council (ERC) under the European Union Horizon 2020 Research and Innovation Programme, Grant Agreement No. 678862 (A. C. B. and M. S. R.). A. C. B. and M. S. R. are also grateful to the Villum Foundation for their support. We also acknowledge funding from DFG (German Research Foundation) under project number 277101999 TRR 183.
\end{acknowledgments}

\bibliography{biblio_ti}

\begin{thebibliography}{57}
\expandafter\ifx\csname natexlab\endcsname\relax\def\natexlab#1{#1}\fi
\expandafter\ifx\csname bibnamefont\endcsname\relax
  \def\bibnamefont#1{#1}\fi
\expandafter\ifx\csname bibfnamefont\endcsname\relax
  \def\bibfnamefont#1{#1}\fi
\expandafter\ifx\csname citenamefont\endcsname\relax
  \def\citenamefont#1{#1}\fi
\expandafter\ifx\csname url\endcsname\relax
  \def\url#1{\texttt{#1}}\fi
\expandafter\ifx\csname urlprefix\endcsname\relax\def\urlprefix{URL }\fi
\providecommand{\bibinfo}[2]{#2}
\providecommand{\eprint}[2][]{\url{#2}}

\bibitem[{\citenamefont{Kumai et~al.}(1999)\citenamefont{Kumai, Okimoto, and
  Tokura}}]{Kumai99}
\bibinfo{author}{\bibfnamefont{R.}~\bibnamefont{Kumai}},
  \bibinfo{author}{\bibfnamefont{Y.}~\bibnamefont{Okimoto}}, \bibnamefont{and}
  \bibinfo{author}{\bibfnamefont{Y.}~\bibnamefont{Tokura}},
  \bibinfo{journal}{Science} \textbf{\bibinfo{volume}{284}},
  \bibinfo{pages}{1645} (\bibinfo{year}{1999}).

\bibitem[{\citenamefont{Dalidovich and Phillips}(2004)}]{Dalidovich04}
\bibinfo{author}{\bibfnamefont{D.}~\bibnamefont{Dalidovich}} \bibnamefont{and}
  \bibinfo{author}{\bibfnamefont{P.}~\bibnamefont{Phillips}},
  \bibinfo{journal}{Phys. Rev. Lett.} \textbf{\bibinfo{volume}{93}},
  \bibinfo{pages}{027004} (\bibinfo{year}{2004}).

\bibitem[{\citenamefont{Green and Sondhi}(2005)}]{Green05}
\bibinfo{author}{\bibfnamefont{A.~G.} \bibnamefont{Green}} \bibnamefont{and}
  \bibinfo{author}{\bibfnamefont{S.~L.} \bibnamefont{Sondhi}},
  \bibinfo{journal}{Phys. Rev. Lett.} \textbf{\bibinfo{volume}{95}},
  \bibinfo{pages}{267001} (\bibinfo{year}{2005}).

\bibitem[{\citenamefont{Feldman}(2005)}]{Feldman05}
\bibinfo{author}{\bibfnamefont{D.~E.} \bibnamefont{Feldman}},
  \bibinfo{journal}{Phys. Rev. Lett.} \textbf{\bibinfo{volume}{95}},
  \bibinfo{pages}{177201} (\bibinfo{year}{2005}).

\bibitem[{\citenamefont{Mitra and Millis}(2008)}]{Mitra08}
\bibinfo{author}{\bibfnamefont{A.}~\bibnamefont{Mitra}} \bibnamefont{and}
  \bibinfo{author}{\bibfnamefont{A.~J.} \bibnamefont{Millis}},
  \bibinfo{journal}{Phys. Rev. B} \textbf{\bibinfo{volume}{77}},
  \bibinfo{pages}{220404} (\bibinfo{year}{2008}).

\bibitem[{\citenamefont{Diehl et~al.}(2010)\citenamefont{Diehl, Tomadin,
  Micheli, Fazio, and Zoller}}]{Diehl10}
\bibinfo{author}{\bibfnamefont{S.}~\bibnamefont{Diehl}},
  \bibinfo{author}{\bibfnamefont{A.}~\bibnamefont{Tomadin}},
  \bibinfo{author}{\bibfnamefont{A.}~\bibnamefont{Micheli}},
  \bibinfo{author}{\bibfnamefont{R.}~\bibnamefont{Fazio}}, \bibnamefont{and}
  \bibinfo{author}{\bibfnamefont{P.}~\bibnamefont{Zoller}},
  \bibinfo{journal}{Phys. Rev. Lett.} \textbf{\bibinfo{volume}{105}},
  \bibinfo{pages}{015702} (\bibinfo{year}{2010}).

\bibitem[{\citenamefont{Fausti et~al.}(2011)\citenamefont{Fausti, Tobey, Dean,
  Kaiser, Dienst, Hoffmann, Pyon, Takayama, Takagi, and Cavalleri}}]{Fausti11}
\bibinfo{author}{\bibfnamefont{D.}~\bibnamefont{Fausti}},
  \bibinfo{author}{\bibfnamefont{R.~I.} \bibnamefont{Tobey}},
  \bibinfo{author}{\bibfnamefont{N.}~\bibnamefont{Dean}},
  \bibinfo{author}{\bibfnamefont{S.}~\bibnamefont{Kaiser}},
  \bibinfo{author}{\bibfnamefont{A.}~\bibnamefont{Dienst}},
  \bibinfo{author}{\bibfnamefont{M.~C.} \bibnamefont{Hoffmann}},
  \bibinfo{author}{\bibfnamefont{S.}~\bibnamefont{Pyon}},
  \bibinfo{author}{\bibfnamefont{T.}~\bibnamefont{Takayama}},
  \bibinfo{author}{\bibfnamefont{H.}~\bibnamefont{Takagi}}, \bibnamefont{and}
  \bibinfo{author}{\bibfnamefont{A.}~\bibnamefont{Cavalleri}},
  \bibinfo{journal}{Science} \textbf{\bibinfo{volume}{331}},
  \bibinfo{pages}{189} (\bibinfo{year}{2011}).

\bibitem[{\citenamefont{Rudner and Song}(2019)}]{Rudner18}
\bibinfo{author}{\bibfnamefont{M.~S.} \bibnamefont{Rudner}} \bibnamefont{and}
  \bibinfo{author}{\bibfnamefont{J.~C.~W.} \bibnamefont{Song}},
  \bibinfo{journal}{Nature Physics} \textbf{\bibinfo{volume}{15}},
  \bibinfo{pages}{1017} (\bibinfo{year}{2019}).

\bibitem[{\citenamefont{Basov et~al.}(2017)\citenamefont{Basov, Averitt, and
  Hsieh}}]{Basov17}
\bibinfo{author}{\bibfnamefont{D.~N.} \bibnamefont{Basov}},
  \bibinfo{author}{\bibfnamefont{R.~D.} \bibnamefont{Averitt}},
  \bibnamefont{and} \bibinfo{author}{\bibfnamefont{D.}~\bibnamefont{Hsieh}},
  \bibinfo{journal}{Nature Materials} \textbf{\bibinfo{volume}{16}},
  \bibinfo{pages}{1077} (\bibinfo{year}{2017}).

\bibitem[{\citenamefont{Hasan and Kane}(2010)}]{Hasan10}
\bibinfo{author}{\bibfnamefont{M.~Z.} \bibnamefont{Hasan}} \bibnamefont{and}
  \bibinfo{author}{\bibfnamefont{C.~L.} \bibnamefont{Kane}},
  \bibinfo{journal}{Rev. Mod. Phys.} \textbf{\bibinfo{volume}{82}},
  \bibinfo{pages}{3045} (\bibinfo{year}{2010}).

\bibitem[{\citenamefont{Wu et~al.}(2006)\citenamefont{Wu, Bernevig, and
  Zhang}}]{Wu2006}
\bibinfo{author}{\bibfnamefont{C.}~\bibnamefont{Wu}},
  \bibinfo{author}{\bibfnamefont{B.~A.} \bibnamefont{Bernevig}},
  \bibnamefont{and} \bibinfo{author}{\bibfnamefont{S.-C.} \bibnamefont{Zhang}},
  \bibinfo{journal}{Phys. Rev. Lett.} \textbf{\bibinfo{volume}{96}},
  \bibinfo{pages}{106401} (\bibinfo{year}{2006}).

\bibitem[{\citenamefont{Xu and Moore}(2006)}]{Xu2006}
\bibinfo{author}{\bibfnamefont{C.}~\bibnamefont{Xu}} \bibnamefont{and}
  \bibinfo{author}{\bibfnamefont{J.~E.} \bibnamefont{Moore}},
  \bibinfo{journal}{Phys. Rev. B} \textbf{\bibinfo{volume}{73}},
  \bibinfo{pages}{045322} (\bibinfo{year}{2006}).

\bibitem[{\citenamefont{Baum and Stern}(2012)}]{Baum12}
\bibinfo{author}{\bibfnamefont{Y.}~\bibnamefont{Baum}} \bibnamefont{and}
  \bibinfo{author}{\bibfnamefont{A.}~\bibnamefont{Stern}},
  \bibinfo{journal}{Phys. Rev. B} \textbf{\bibinfo{volume}{85}},
  \bibinfo{pages}{121105} (\bibinfo{year}{2012}).

\bibitem[{\citenamefont{Black-Schaffer and Yudin}(2014)}]{Black-Schaffer14}
\bibinfo{author}{\bibfnamefont{A.~M.} \bibnamefont{Black-Schaffer}}
  \bibnamefont{and} \bibinfo{author}{\bibfnamefont{D.}~\bibnamefont{Yudin}},
  \bibinfo{journal}{Phys. Rev. B} \textbf{\bibinfo{volume}{90}},
  \bibinfo{pages}{161413} (\bibinfo{year}{2014}).

\bibitem[{\citenamefont{Chou et~al.}(2018)\citenamefont{Chou, Nandkishore, and
  Radzihovsky}}]{Chou18}
\bibinfo{author}{\bibfnamefont{Y.-Z.} \bibnamefont{Chou}},
  \bibinfo{author}{\bibfnamefont{R.~M.} \bibnamefont{Nandkishore}},
  \bibnamefont{and}
  \bibinfo{author}{\bibfnamefont{L.}~\bibnamefont{Radzihovsky}},
  \bibinfo{journal}{Phys. Rev. B} \textbf{\bibinfo{volume}{98}},
  \bibinfo{pages}{054205} (\bibinfo{year}{2018}).

\bibitem[{\citenamefont{Chou et~al.}(2019)\citenamefont{Chou, Nandkishore, and
  Radzihovsky}}]{Chou19}
\bibinfo{author}{\bibfnamefont{Y.-Z.} \bibnamefont{Chou}},
  \bibinfo{author}{\bibfnamefont{R.~M.} \bibnamefont{Nandkishore}},
  \bibnamefont{and}
  \bibinfo{author}{\bibfnamefont{L.}~\bibnamefont{Radzihovsky}},
  \bibinfo{journal}{Phys. Rev. B} \textbf{\bibinfo{volume}{99}},
  \bibinfo{pages}{165108} (\bibinfo{year}{2019}).

\bibitem[{\citenamefont{Liu et~al.}(2008)\citenamefont{Liu, Qi, Dai, Fang, and
  Zhang}}]{Liu08}
\bibinfo{author}{\bibfnamefont{C.-X.} \bibnamefont{Liu}},
  \bibinfo{author}{\bibfnamefont{X.-L.} \bibnamefont{Qi}},
  \bibinfo{author}{\bibfnamefont{X.}~\bibnamefont{Dai}},
  \bibinfo{author}{\bibfnamefont{Z.}~\bibnamefont{Fang}}, \bibnamefont{and}
  \bibinfo{author}{\bibfnamefont{S.-C.} \bibnamefont{Zhang}},
  \bibinfo{journal}{Phys. Rev. Lett.} \textbf{\bibinfo{volume}{101}},
  \bibinfo{pages}{146802} (\bibinfo{year}{2008}).

\bibitem[{\citenamefont{Chen et~al.}(2010)\citenamefont{Chen, Chu, Analytis,
  Liu, Igarashi, Kuo, Qi, Mo, Moore, Lu et~al.}}]{Chen10}
\bibinfo{author}{\bibfnamefont{Y.~L.} \bibnamefont{Chen}},
  \bibinfo{author}{\bibfnamefont{J.-H.} \bibnamefont{Chu}},
  \bibinfo{author}{\bibfnamefont{J.~G.} \bibnamefont{Analytis}},
  \bibinfo{author}{\bibfnamefont{Z.~K.} \bibnamefont{Liu}},
  \bibinfo{author}{\bibfnamefont{K.}~\bibnamefont{Igarashi}},
  \bibinfo{author}{\bibfnamefont{H.-H.} \bibnamefont{Kuo}},
  \bibinfo{author}{\bibfnamefont{X.~L.} \bibnamefont{Qi}},
  \bibinfo{author}{\bibfnamefont{S.~K.} \bibnamefont{Mo}},
  \bibinfo{author}{\bibfnamefont{R.~G.} \bibnamefont{Moore}},
  \bibinfo{author}{\bibfnamefont{D.~H.} \bibnamefont{Lu}},
  \bibnamefont{et~al.}, \bibinfo{journal}{Science}
  \textbf{\bibinfo{volume}{329}}, \bibinfo{pages}{659} (\bibinfo{year}{2010}).

\bibitem[{\citenamefont{Yu et~al.}(2010)\citenamefont{Yu, Zhang, Zhang, Zhang,
  Dai, and Fang}}]{Yu10}
\bibinfo{author}{\bibfnamefont{R.}~\bibnamefont{Yu}},
  \bibinfo{author}{\bibfnamefont{W.}~\bibnamefont{Zhang}},
  \bibinfo{author}{\bibfnamefont{H.-J.} \bibnamefont{Zhang}},
  \bibinfo{author}{\bibfnamefont{S.-C.} \bibnamefont{Zhang}},
  \bibinfo{author}{\bibfnamefont{X.}~\bibnamefont{Dai}}, \bibnamefont{and}
  \bibinfo{author}{\bibfnamefont{Z.}~\bibnamefont{Fang}},
  \bibinfo{journal}{Science} \textbf{\bibinfo{volume}{329}},
  \bibinfo{pages}{61} (\bibinfo{year}{2010}).

\bibitem[{\citenamefont{Zhu et~al.}(2011)\citenamefont{Zhu, Yao, Zhang, and
  Chang}}]{Zhu11}
\bibinfo{author}{\bibfnamefont{J.-J.} \bibnamefont{Zhu}},
  \bibinfo{author}{\bibfnamefont{D.-X.} \bibnamefont{Yao}},
  \bibinfo{author}{\bibfnamefont{S.-C.} \bibnamefont{Zhang}}, \bibnamefont{and}
  \bibinfo{author}{\bibfnamefont{K.}~\bibnamefont{Chang}},
  \bibinfo{journal}{Phys. Rev. Lett.} \textbf{\bibinfo{volume}{106}},
  \bibinfo{pages}{097201} (\bibinfo{year}{2011}).

\bibitem[{\citenamefont{Fox et~al.}(2018)\citenamefont{Fox, Rosen, Yang, Jones,
  Elmquist, Kou, Pan, Wang, and Goldhaber-Gordon}}]{Fox18}
\bibinfo{author}{\bibfnamefont{E.~J.} \bibnamefont{Fox}},
  \bibinfo{author}{\bibfnamefont{I.~T.} \bibnamefont{Rosen}},
  \bibinfo{author}{\bibfnamefont{Y.}~\bibnamefont{Yang}},
  \bibinfo{author}{\bibfnamefont{G.~R.} \bibnamefont{Jones}},
  \bibinfo{author}{\bibfnamefont{R.~E.} \bibnamefont{Elmquist}},
  \bibinfo{author}{\bibfnamefont{X.}~\bibnamefont{Kou}},
  \bibinfo{author}{\bibfnamefont{L.}~\bibnamefont{Pan}},
  \bibinfo{author}{\bibfnamefont{K.~L.} \bibnamefont{Wang}}, \bibnamefont{and}
  \bibinfo{author}{\bibfnamefont{D.}~\bibnamefont{Goldhaber-Gordon}},
  \bibinfo{journal}{Phys. Rev. B} \textbf{\bibinfo{volume}{98}},
  \bibinfo{pages}{075145} (\bibinfo{year}{2018}).

\bibitem[{\citenamefont{Tang et~al.}(2017)\citenamefont{Tang, Chang, Zhao, Liu,
  Jiang, Liu, McCartney, Smith, Chen, Moodera et~al.}}]{Tang17}
\bibinfo{author}{\bibfnamefont{C.}~\bibnamefont{Tang}},
  \bibinfo{author}{\bibfnamefont{C.-Z.} \bibnamefont{Chang}},
  \bibinfo{author}{\bibfnamefont{G.}~\bibnamefont{Zhao}},
  \bibinfo{author}{\bibfnamefont{Y.}~\bibnamefont{Liu}},
  \bibinfo{author}{\bibfnamefont{Z.}~\bibnamefont{Jiang}},
  \bibinfo{author}{\bibfnamefont{C.-X.} \bibnamefont{Liu}},
  \bibinfo{author}{\bibfnamefont{M.~R.} \bibnamefont{McCartney}},
  \bibinfo{author}{\bibfnamefont{D.~J.} \bibnamefont{Smith}},
  \bibinfo{author}{\bibfnamefont{T.}~\bibnamefont{Chen}},
  \bibinfo{author}{\bibfnamefont{J.~S.} \bibnamefont{Moodera}},
  \bibnamefont{et~al.}, \bibinfo{journal}{Science Advances}
  \textbf{\bibinfo{volume}{3}}, \bibinfo{pages}{e1700307}
  (\bibinfo{year}{2017}).

\bibitem[{\citenamefont{Del~Maestro et~al.}(2013)\citenamefont{Del~Maestro,
  Hyart, and Rosenow}}]{Maestro13}
\bibinfo{author}{\bibfnamefont{A.}~\bibnamefont{Del~Maestro}},
  \bibinfo{author}{\bibfnamefont{T.}~\bibnamefont{Hyart}}, \bibnamefont{and}
  \bibinfo{author}{\bibfnamefont{B.}~\bibnamefont{Rosenow}},
  \bibinfo{journal}{Phys. Rev. B} \textbf{\bibinfo{volume}{87}},
  \bibinfo{pages}{165440} (\bibinfo{year}{2013}).

\bibitem[{\citenamefont{{Hsu, Chen-Hsuan and Stano, Peter and Klinovaja, Jelena
  and Loss, Daniel}}(2017)}]{Hsu17}
\bibinfo{author}{\bibnamefont{{Hsu, Chen-Hsuan and Stano, Peter and Klinovaja,
  Jelena and Loss, Daniel}}}, \bibinfo{journal}{Phys. Rev. B}
  \textbf{\bibinfo{volume}{96}}, \bibinfo{pages}{081405}
  (\bibinfo{year}{2017}).

\bibitem[{\citenamefont{{Hsu, Chen-Hsuan and Stano, Peter and Klinovaja, Jelena
  and Loss, Daniel}}(2018)}]{Hsu18}
\bibinfo{author}{\bibnamefont{{Hsu, Chen-Hsuan and Stano, Peter and Klinovaja,
  Jelena and Loss, Daniel}}}, \bibinfo{journal}{Phys. Rev. B}
  \textbf{\bibinfo{volume}{97}}, \bibinfo{pages}{125432}
  (\bibinfo{year}{2018}).

\bibitem[{\citenamefont{K{\"o}nig et~al.}(2007)\citenamefont{K{\"o}nig,
  Wiedmann, Br{\"u}ne, Roth, Buhmann, Molenkamp, Qi, and Zhang}}]{Konig07}
\bibinfo{author}{\bibfnamefont{M.}~\bibnamefont{K{\"o}nig}},
  \bibinfo{author}{\bibfnamefont{S.}~\bibnamefont{Wiedmann}},
  \bibinfo{author}{\bibfnamefont{C.}~\bibnamefont{Br{\"u}ne}},
  \bibinfo{author}{\bibfnamefont{A.}~\bibnamefont{Roth}},
  \bibinfo{author}{\bibfnamefont{H.}~\bibnamefont{Buhmann}},
  \bibinfo{author}{\bibfnamefont{L.~W.} \bibnamefont{Molenkamp}},
  \bibinfo{author}{\bibfnamefont{X.-L.} \bibnamefont{Qi}}, \bibnamefont{and}
  \bibinfo{author}{\bibfnamefont{S.-C.} \bibnamefont{Zhang}},
  \bibinfo{journal}{Science} \textbf{\bibinfo{volume}{318}},
  \bibinfo{pages}{766} (\bibinfo{year}{2007}).

\bibitem[{\citenamefont{Dominguez et~al.}(2018)\citenamefont{Dominguez, Scharf,
  Li, Sch\"afer, Claessen, Hanke, Thomale, and Hankiewicz}}]{Dominguez18}
\bibinfo{author}{\bibfnamefont{F.}~\bibnamefont{Dominguez}},
  \bibinfo{author}{\bibfnamefont{B.}~\bibnamefont{Scharf}},
  \bibinfo{author}{\bibfnamefont{G.}~\bibnamefont{Li}},
  \bibinfo{author}{\bibfnamefont{J.}~\bibnamefont{Sch\"afer}},
  \bibinfo{author}{\bibfnamefont{R.}~\bibnamefont{Claessen}},
  \bibinfo{author}{\bibfnamefont{W.}~\bibnamefont{Hanke}},
  \bibinfo{author}{\bibfnamefont{R.}~\bibnamefont{Thomale}}, \bibnamefont{and}
  \bibinfo{author}{\bibfnamefont{E.~M.} \bibnamefont{Hankiewicz}},
  \bibinfo{journal}{Phys. Rev. B} \textbf{\bibinfo{volume}{98}},
  \bibinfo{pages}{161407} (\bibinfo{year}{2018}).

\bibitem[{\citenamefont{Skolasinski et~al.}(2018)\citenamefont{Skolasinski,
  Pikulin, Alicea, and Wimmer}}]{Skolasinski18}
\bibinfo{author}{\bibfnamefont{R.}~\bibnamefont{Skolasinski}},
  \bibinfo{author}{\bibfnamefont{D.~I.} \bibnamefont{Pikulin}},
  \bibinfo{author}{\bibfnamefont{J.}~\bibnamefont{Alicea}}, \bibnamefont{and}
  \bibinfo{author}{\bibfnamefont{M.}~\bibnamefont{Wimmer}},
  \bibinfo{journal}{Phys. Rev. B} \textbf{\bibinfo{volume}{98}},
  \bibinfo{pages}{201404} (\bibinfo{year}{2018}).

\bibitem[{\citenamefont{Wang et~al.}(2013)\citenamefont{Wang, Steinberg,
  Jarillo-Herrero, and Gedik}}]{Wang13}
\bibinfo{author}{\bibfnamefont{Y.~H.} \bibnamefont{Wang}},
  \bibinfo{author}{\bibfnamefont{H.}~\bibnamefont{Steinberg}},
  \bibinfo{author}{\bibfnamefont{P.}~\bibnamefont{Jarillo-Herrero}},
  \bibnamefont{and} \bibinfo{author}{\bibfnamefont{N.}~\bibnamefont{Gedik}},
  \bibinfo{journal}{Science} \textbf{\bibinfo{volume}{342}},
  \bibinfo{pages}{453} (\bibinfo{year}{2013}).

\bibitem[{\citenamefont{Edelstein}(1990)}]{Edelstein90}
\bibinfo{author}{\bibfnamefont{V.}~\bibnamefont{Edelstein}},
  \bibinfo{journal}{Solid State Communications} \textbf{\bibinfo{volume}{73}},
  \bibinfo{pages}{233 } (\bibinfo{year}{1990}).

\bibitem[{\citenamefont{Br{\"u}ne et~al.}(2012)\citenamefont{Br{\"u}ne, Roth,
  Buhmann, Hankiewicz, Molenkamp, Maciejko, Qi, and Zhang}}]{Brune12}
\bibinfo{author}{\bibfnamefont{C.}~\bibnamefont{Br{\"u}ne}},
  \bibinfo{author}{\bibfnamefont{A.}~\bibnamefont{Roth}},
  \bibinfo{author}{\bibfnamefont{H.}~\bibnamefont{Buhmann}},
  \bibinfo{author}{\bibfnamefont{E.~M.} \bibnamefont{Hankiewicz}},
  \bibinfo{author}{\bibfnamefont{L.~W.} \bibnamefont{Molenkamp}},
  \bibinfo{author}{\bibfnamefont{J.}~\bibnamefont{Maciejko}},
  \bibinfo{author}{\bibfnamefont{X.-L.} \bibnamefont{Qi}}, \bibnamefont{and}
  \bibinfo{author}{\bibfnamefont{S.-C.} \bibnamefont{Zhang}},
  \bibinfo{journal}{Nature Physics} \textbf{\bibinfo{volume}{8}},
  \bibinfo{pages}{485 EP } (\bibinfo{year}{2012}), \bibinfo{note}{article}.

\bibitem[{\citenamefont{Schmidt et~al.}(2012)\citenamefont{Schmidt, Rachel, von
  Oppen, and Glazman}}]{Schmidt12}
\bibinfo{author}{\bibfnamefont{T.~L.} \bibnamefont{Schmidt}},
  \bibinfo{author}{\bibfnamefont{S.}~\bibnamefont{Rachel}},
  \bibinfo{author}{\bibfnamefont{F.}~\bibnamefont{von Oppen}},
  \bibnamefont{and} \bibinfo{author}{\bibfnamefont{L.~I.}
  \bibnamefont{Glazman}}, \bibinfo{journal}{Phys. Rev. Lett.}
  \textbf{\bibinfo{volume}{108}}, \bibinfo{pages}{156402}
  (\bibinfo{year}{2012}).

\bibitem[{\citenamefont{Rod et~al.}(2015)\citenamefont{Rod, Schmidt, and
  Rachel}}]{Rod15}
\bibinfo{author}{\bibfnamefont{A.}~\bibnamefont{Rod}},
  \bibinfo{author}{\bibfnamefont{T.~L.} \bibnamefont{Schmidt}},
  \bibnamefont{and} \bibinfo{author}{\bibfnamefont{S.}~\bibnamefont{Rachel}},
  \bibinfo{journal}{Phys. Rev. B} \textbf{\bibinfo{volume}{91}},
  \bibinfo{pages}{245112} (\bibinfo{year}{2015}).

\bibitem[{\citenamefont{Ortiz et~al.}(2016)\citenamefont{Ortiz, Molina,
  Platero, and Lunde}}]{Ortiz16}
\bibinfo{author}{\bibfnamefont{L.}~\bibnamefont{Ortiz}},
  \bibinfo{author}{\bibfnamefont{R.~A.} \bibnamefont{Molina}},
  \bibinfo{author}{\bibfnamefont{G.}~\bibnamefont{Platero}}, \bibnamefont{and}
  \bibinfo{author}{\bibfnamefont{A.~M.} \bibnamefont{Lunde}},
  \bibinfo{journal}{Phys. Rev. B} \textbf{\bibinfo{volume}{93}},
  \bibinfo{pages}{205431} (\bibinfo{year}{2016}).

\bibitem[{foo({\natexlab{a}})}]{footnote:no_backscattering}
\bibinfo{note}{We note that TRS in {\it nonhelical} spin-orbit coupled systems
  (i.e., systems that are not realized as surface states of higher-dimensional
  topological insluators) does {\it not} provide protection against elastic
  backscattering.}

\bibitem[{\citenamefont{Chang and Lou}(2011)}]{Chang11}
\bibinfo{author}{\bibfnamefont{K.}~\bibnamefont{Chang}} \bibnamefont{and}
  \bibinfo{author}{\bibfnamefont{W.-K.} \bibnamefont{Lou}},
  \bibinfo{journal}{Phys. Rev. Lett.} \textbf{\bibinfo{volume}{106}},
  \bibinfo{pages}{206802} (\bibinfo{year}{2011}).

\bibitem[{foo({\natexlab{b}})}]{footnote:k3sz}
\bibinfo{note}{A spin-independent term proportional to $k^2$ and a term
  proportional to $k^3 \sigma_z$ would also be allowed by symmetry. However,
  the presence of such terms would not alter the qualitative features of our
  work and therefore we omit them for simplicity and clarity.}

\bibitem[{foo({\natexlab{c}})}]{footnote:self_consistency}
\bibinfo{note}{In the perturbative regime that we consider, a self-consistent
  solution for the current-induced spin polarization, calculated using the mean
  field band structure from
  Eq.~(\ref{eq_Hamiltonian_QSH_edge_cubic_SOC_driven_interacting}), yields only
  minor quantitative corrections to our results.}

\bibitem[{foo({\natexlab{d}})}]{footnote:velocity}
\bibinfo{note}{Within the limits that we take, the dispersion is effectively
  linearized, $\varepsilon(k) \approx \hbar v k$, while the spin helicity axis
  rotation is retained. Accounting for curvature of the dispersion may
  introduce small quantitative corrections, but it will not change the
  qualitative behavior.}

\bibitem[{\citenamefont{V\"ayrynen et~al.}(2013)\citenamefont{V\"ayrynen,
  Goldstein, and Glazman}}]{Vayrynen2013}
\bibinfo{author}{\bibfnamefont{J.~I.} \bibnamefont{V\"ayrynen}},
  \bibinfo{author}{\bibfnamefont{M.}~\bibnamefont{Goldstein}},
  \bibnamefont{and} \bibinfo{author}{\bibfnamefont{L.~I.}
  \bibnamefont{Glazman}}, \bibinfo{journal}{Phys. Rev. Lett.}
  \textbf{\bibinfo{volume}{110}}, \bibinfo{pages}{216402}
  (\bibinfo{year}{2013}).

\bibitem[{\citenamefont{Bagrov et~al.}(2018)\citenamefont{Bagrov, Guinea, and
  Katsnelson}}]{Bagrov2018}
\bibinfo{author}{\bibfnamefont{A.~A.} \bibnamefont{Bagrov}},
  \bibinfo{author}{\bibfnamefont{F.}~\bibnamefont{Guinea}}, \bibnamefont{and}
  \bibinfo{author}{\bibfnamefont{M.~I.} \bibnamefont{Katsnelson}},
  \bibinfo{journal}{arXiv:1805.11700}  (\bibinfo{year}{2018}).

\bibitem[{\citenamefont{Novelli et~al.}(2019)\citenamefont{Novelli, Taddei,
  Geim, and Polini}}]{Novelli2019}
\bibinfo{author}{\bibfnamefont{P.}~\bibnamefont{Novelli}},
  \bibinfo{author}{\bibfnamefont{F.}~\bibnamefont{Taddei}},
  \bibinfo{author}{\bibfnamefont{A.~K.} \bibnamefont{Geim}}, \bibnamefont{and}
  \bibinfo{author}{\bibfnamefont{M.}~\bibnamefont{Polini}},
  \bibinfo{journal}{Phys. Rev. Lett.} \textbf{\bibinfo{volume}{122}},
  \bibinfo{pages}{016601} (\bibinfo{year}{2019}).

\bibitem[{\citenamefont{Fu}(2009)}]{Fu09}
\bibinfo{author}{\bibfnamefont{L.}~\bibnamefont{Fu}}, \bibinfo{journal}{Phys.
  Rev. Lett.} \textbf{\bibinfo{volume}{103}}, \bibinfo{pages}{266801}
  (\bibinfo{year}{2009}).

\bibitem[{\citenamefont{Lee et~al.}(2009)\citenamefont{Lee, Wu, Arovas, and
  Zhang}}]{Lee09}
\bibinfo{author}{\bibfnamefont{W.-C.} \bibnamefont{Lee}},
  \bibinfo{author}{\bibfnamefont{C.}~\bibnamefont{Wu}},
  \bibinfo{author}{\bibfnamefont{D.~P.} \bibnamefont{Arovas}},
  \bibnamefont{and} \bibinfo{author}{\bibfnamefont{S.-C.} \bibnamefont{Zhang}},
  \bibinfo{journal}{Phys. Rev. B} \textbf{\bibinfo{volume}{80}},
  \bibinfo{pages}{245439} (\bibinfo{year}{2009}).

\bibitem[{\citenamefont{Sodemann and Fu}(2015)}]{Sodemann15}
\bibinfo{author}{\bibfnamefont{I.}~\bibnamefont{Sodemann}} \bibnamefont{and}
  \bibinfo{author}{\bibfnamefont{L.}~\bibnamefont{Fu}}, \bibinfo{journal}{Phys.
  Rev. Lett.} \textbf{\bibinfo{volume}{115}}, \bibinfo{pages}{216806}
  (\bibinfo{year}{2015}).

\bibitem[{\citenamefont{Qian et~al.}(2014)\citenamefont{Qian, Liu, Fu, and
  Li}}]{Qian14}
\bibinfo{author}{\bibfnamefont{X.}~\bibnamefont{Qian}},
  \bibinfo{author}{\bibfnamefont{J.}~\bibnamefont{Liu}},
  \bibinfo{author}{\bibfnamefont{L.}~\bibnamefont{Fu}}, \bibnamefont{and}
  \bibinfo{author}{\bibfnamefont{J.}~\bibnamefont{Li}},
  \bibinfo{journal}{Science} \textbf{\bibinfo{volume}{346}},
  \bibinfo{pages}{1344} (\bibinfo{year}{2014}).

\bibitem[{\citenamefont{Fei et~al.}(2017)\citenamefont{Fei, Palomaki, Wu, Zhao,
  Cai, Sun, Nguyen, Finney, Xu, and Cobden}}]{Fei17}
\bibinfo{author}{\bibfnamefont{Z.}~\bibnamefont{Fei}},
  \bibinfo{author}{\bibfnamefont{T.}~\bibnamefont{Palomaki}},
  \bibinfo{author}{\bibfnamefont{S.}~\bibnamefont{Wu}},
  \bibinfo{author}{\bibfnamefont{W.}~\bibnamefont{Zhao}},
  \bibinfo{author}{\bibfnamefont{X.}~\bibnamefont{Cai}},
  \bibinfo{author}{\bibfnamefont{B.}~\bibnamefont{Sun}},
  \bibinfo{author}{\bibfnamefont{P.}~\bibnamefont{Nguyen}},
  \bibinfo{author}{\bibfnamefont{J.}~\bibnamefont{Finney}},
  \bibinfo{author}{\bibfnamefont{X.}~\bibnamefont{Xu}}, \bibnamefont{and}
  \bibinfo{author}{\bibfnamefont{D.~H.} \bibnamefont{Cobden}},
  \bibinfo{journal}{Nature Physics} \textbf{\bibinfo{volume}{13}},
  \bibinfo{pages}{677} (\bibinfo{year}{2017}).

\bibitem[{\citenamefont{Wu et~al.}(2018)\citenamefont{Wu, Fatemi, Gibson,
  Watanabe, Taniguchi, Cava, and Jarillo-Herrero}}]{Wu18}
\bibinfo{author}{\bibfnamefont{A.}~\bibnamefont{Wu}},
  \bibinfo{author}{\bibfnamefont{V.}~\bibnamefont{Fatemi}},
  \bibinfo{author}{\bibfnamefont{Q.~D.} \bibnamefont{Gibson}},
  \bibinfo{author}{\bibfnamefont{K.}~\bibnamefont{Watanabe}},
  \bibinfo{author}{\bibfnamefont{T.}~\bibnamefont{Taniguchi}},
  \bibinfo{author}{\bibfnamefont{R.~J.} \bibnamefont{Cava}}, \bibnamefont{and}
  \bibinfo{author}{\bibfnamefont{P.}~\bibnamefont{Jarillo-Herrero}},
  \bibinfo{journal}{Science} \textbf{\bibinfo{volume}{359}},
  \bibinfo{pages}{76} (\bibinfo{year}{2018}).

\bibitem[{\citenamefont{Marrazzo et~al.}(2018)\citenamefont{Marrazzo,
  Gibertini, Campi, Mounet, and Marzari}}]{Marrazzo18}
\bibinfo{author}{\bibfnamefont{A.}~\bibnamefont{Marrazzo}},
  \bibinfo{author}{\bibfnamefont{M.}~\bibnamefont{Gibertini}},
  \bibinfo{author}{\bibfnamefont{D.}~\bibnamefont{Campi}},
  \bibinfo{author}{\bibfnamefont{N.}~\bibnamefont{Mounet}}, \bibnamefont{and}
  \bibinfo{author}{\bibfnamefont{N.}~\bibnamefont{Marzari}},
  \bibinfo{journal}{Phys. Rev. Lett.} \textbf{\bibinfo{volume}{120}},
  \bibinfo{pages}{117701} (\bibinfo{year}{2018}).

\bibitem[{\citenamefont{Xu et~al.}(2013)\citenamefont{Xu, Yan, Zhang, Wang, Xu,
  Tang, Duan, and Zhang}}]{Xu13}
\bibinfo{author}{\bibfnamefont{Y.}~\bibnamefont{Xu}},
  \bibinfo{author}{\bibfnamefont{B.}~\bibnamefont{Yan}},
  \bibinfo{author}{\bibfnamefont{H.-J.} \bibnamefont{Zhang}},
  \bibinfo{author}{\bibfnamefont{J.}~\bibnamefont{Wang}},
  \bibinfo{author}{\bibfnamefont{G.}~\bibnamefont{Xu}},
  \bibinfo{author}{\bibfnamefont{P.}~\bibnamefont{Tang}},
  \bibinfo{author}{\bibfnamefont{W.}~\bibnamefont{Duan}}, \bibnamefont{and}
  \bibinfo{author}{\bibfnamefont{S.-C.} \bibnamefont{Zhang}},
  \bibinfo{journal}{Phys. Rev. Lett.} \textbf{\bibinfo{volume}{111}},
  \bibinfo{pages}{136804} (\bibinfo{year}{2013}).

\bibitem[{foo({\natexlab{e}})}]{footnote:g_value}
\bibinfo{note}{We note that this value of $g$ corresponds to a Luttinger
  parameter larger than the threshold value for the breakdown of edge states
  found in Ref.~\cite{Wu2006}.}

\bibitem[{\citenamefont{Laturia et~al.}(2018)\citenamefont{Laturia, Van~de Put,
  and Vandenberghe}}]{Laturia18}
\bibinfo{author}{\bibfnamefont{A.}~\bibnamefont{Laturia}},
  \bibinfo{author}{\bibfnamefont{M.~L.} \bibnamefont{Van~de Put}},
  \bibnamefont{and} \bibinfo{author}{\bibfnamefont{W.~G.}
  \bibnamefont{Vandenberghe}}, \bibinfo{journal}{npj 2D Materials and
  Applications} \textbf{\bibinfo{volume}{2}}, \bibinfo{pages}{6}
  (\bibinfo{year}{2018}).

\bibitem[{\citenamefont{{Ma} et~al.}(2019)\citenamefont{{Ma}, {Xu}, {Shen},
  {Macneill}, {Fatemi}, {Mier Valdivia}, {Wu}, {Chang}, {Du}, {Hsu}
  et~al.}}]{Ma18}
\bibinfo{author}{\bibfnamefont{Q.}~\bibnamefont{{Ma}}},
  \bibinfo{author}{\bibfnamefont{S.-Y.} \bibnamefont{{Xu}}},
  \bibinfo{author}{\bibfnamefont{H.}~\bibnamefont{{Shen}}},
  \bibinfo{author}{\bibfnamefont{D.}~\bibnamefont{{Macneill}}},
  \bibinfo{author}{\bibfnamefont{V.}~\bibnamefont{{Fatemi}}},
  \bibinfo{author}{\bibfnamefont{A.~M.} \bibnamefont{{Mier Valdivia}}},
  \bibinfo{author}{\bibfnamefont{S.}~\bibnamefont{{Wu}}},
  \bibinfo{author}{\bibfnamefont{T.-R.} \bibnamefont{{Chang}}},
  \bibinfo{author}{\bibfnamefont{Z.}~\bibnamefont{{Du}}},
  \bibinfo{author}{\bibfnamefont{C.-H.} \bibnamefont{{Hsu}}},
  \bibnamefont{et~al.}, \bibinfo{journal}{Nature}
  \textbf{\bibinfo{volume}{565}}, \bibinfo{pages}{337} (\bibinfo{year}{2019}).

\bibitem[{\citenamefont{Xu et~al.}(2018)\citenamefont{Xu, Ma, Shen, Fatemi, Wu,
  Chang, Chang, Valdivia, Chan, Gibson et~al.}}]{Xu18}
\bibinfo{author}{\bibfnamefont{S.-Y.} \bibnamefont{Xu}},
  \bibinfo{author}{\bibfnamefont{Q.}~\bibnamefont{Ma}},
  \bibinfo{author}{\bibfnamefont{H.}~\bibnamefont{Shen}},
  \bibinfo{author}{\bibfnamefont{V.}~\bibnamefont{Fatemi}},
  \bibinfo{author}{\bibfnamefont{S.}~\bibnamefont{Wu}},
  \bibinfo{author}{\bibfnamefont{T.-R.} \bibnamefont{Chang}},
  \bibinfo{author}{\bibfnamefont{G.}~\bibnamefont{Chang}},
  \bibinfo{author}{\bibfnamefont{A.~M.~M.} \bibnamefont{Valdivia}},
  \bibinfo{author}{\bibfnamefont{C.-K.} \bibnamefont{Chan}},
  \bibinfo{author}{\bibfnamefont{Q.~D.} \bibnamefont{Gibson}},
  \bibnamefont{et~al.}, \bibinfo{journal}{Nature Physics}
  \textbf{\bibinfo{volume}{14}}, \bibinfo{pages}{900} (\bibinfo{year}{2018}).

\bibitem[{\citenamefont{You et~al.}(2018)\citenamefont{You, Fang, Xu, Kaxiras,
  and Low}}]{You18}
\bibinfo{author}{\bibfnamefont{J.-S.} \bibnamefont{You}},
  \bibinfo{author}{\bibfnamefont{S.}~\bibnamefont{Fang}},
  \bibinfo{author}{\bibfnamefont{S.-Y.} \bibnamefont{Xu}},
  \bibinfo{author}{\bibfnamefont{E.}~\bibnamefont{Kaxiras}}, \bibnamefont{and}
  \bibinfo{author}{\bibfnamefont{T.}~\bibnamefont{Low}},
  \bibinfo{journal}{Phys. Rev. B} \textbf{\bibinfo{volume}{98}},
  \bibinfo{pages}{121109} (\bibinfo{year}{2018}).

\bibitem[{\citenamefont{Kang et~al.}(2019)\citenamefont{Kang, Li, Sohn, Shan,
  and Mak}}]{Kang18}
\bibinfo{author}{\bibfnamefont{K.}~\bibnamefont{Kang}},
  \bibinfo{author}{\bibfnamefont{T.}~\bibnamefont{Li}},
  \bibinfo{author}{\bibfnamefont{E.}~\bibnamefont{Sohn}},
  \bibinfo{author}{\bibfnamefont{J.}~\bibnamefont{Shan}}, \bibnamefont{and}
  \bibinfo{author}{\bibfnamefont{K.~F.} \bibnamefont{Mak}},
  \bibinfo{journal}{Nature Materials} \textbf{\bibinfo{volume}{18}},
  \bibinfo{pages}{324} (\bibinfo{year}{2019}).

\bibitem[{\citenamefont{K\"onig et~al.}(2019)\citenamefont{K\"onig, Dzero,
  Levchenko, and Pesin}}]{Konig18}
\bibinfo{author}{\bibfnamefont{E.~J.} \bibnamefont{K\"onig}},
  \bibinfo{author}{\bibfnamefont{M.}~\bibnamefont{Dzero}},
  \bibinfo{author}{\bibfnamefont{A.}~\bibnamefont{Levchenko}},
  \bibnamefont{and} \bibinfo{author}{\bibfnamefont{D.~A.} \bibnamefont{Pesin}},
  \bibinfo{journal}{Phys. Rev. B} \textbf{\bibinfo{volume}{99}},
  \bibinfo{pages}{155404} (\bibinfo{year}{2019}).

\end{thebibliography}
\bibliographystyle{apsrev_nourl}

\end{document}